\documentclass[sigchi-a]{acmart}% draft
\usepackage{graphicx}%
\usepackage{lipsum}%
\usepackage{paralist}% inline lists
\usepackage{color}%
\usepackage[utf8]{inputenc}%

\usepackage[subtle]{savetrees}%

\copyrightyear{2019}
\acmYear{2019}
\setcopyright{rightsretained}
\acmConference[DC\textsuperscript{2}S\textsuperscript{2} '19: Workshop on Designing Crowd-powered Creativity Support Systems]{DC\textsuperscript{2}S\textsuperscript{2} '19: Workshop on Designing Crowd-powered Creativity Support Systems}{May 04--09, 2019}{Glasgow, Scotland UK}
\acmDOI{}
\acmISBN{}

\begin{document}

\title{Supporting Creative~Work with Crowd Feedback Systems}

% The "author" command and its associated commands are used to define the authors and their affiliations.
% Of note is the shared affiliation of the first two authors, and the "authornote" and "authornotemark" commands
% used to denote shared contribution to the research.
\author{Jonas Oppenlaender}
\email{jonas.oppenlaender@oulu.fi}
\orcid{0000-0002-2342-1540}
\affiliation{%
  \institution{Center for Ubiquitous Computing}
  \institution{University of Oulu}
  \city{Oulu}
  \country{Finland}
}

\author{Simo Hosio}
\email{simo.hosio@oulu.fi}
\affiliation{%
  \institution{Center for Ubiquitous Computing}
  \institution{University of Oulu}
  \city{Oulu}
  \country{Finland}
}

% By default, the full list of authors will be used in the page headers. Often, this list is too long, and will overlap
% other information printed in the page headers. This command allows the author to define a more concise list
% of authors' names for this purpose.
\renewcommand{\shortauthors}{Oppenlaender, et al.}

% The abstract is a short summary of the work to be presented in the article.
\begin{abstract}
Crowd feedback systems have the potential to support creative workers with feedback from the crowd.
In this position paper for the \textit{Workshop on Designing Crowd-powered Creativity Support Systems ($DC^2S^2$)} at CHI'19~\cite{dc2s22019},
we present three creativity support tools
% developed at our laboratory
in which we explore how creative workers can be assisted with crowdsourced formative and summative feedback.
% Future work will develop a framework for the design of crowd feedback systems that will specifically highlight crowdsourced feedback that goes beyond mere text-based feedback.
For each of the three crowd feedback systems, we provide one idea for future research.
\end{abstract}




\settopmatter{printacmref=false}

% A "teaser" image appears between the author and affiliation information and the body of the document, and typically spans the page. 
%%\begin{teaserfigure}
%%  \includegraphics[width=\textwidth]{sampleteaser}
%%  \caption{Seattle Mariners at Spring Training, 2010.}
%%  \Description{Enjoying the baseball game from the third-base seats. Ichiro Suzuki preparing to bat.}
%%  \label{fig:teaser}
%%\end{teaserfigure}

\maketitle

\section{Introduction}

Supporting human creativity has been considered as one of the grand challenges of Human-Com\-put\-er Interaction  (HCI)~\cite{Shneiderman2009}. % shneidermanWorkshop
Crowdsourcing~\cite{howe2006} is a powerful approach for tapping into the collective insights of a diverse crowd of people.
The combination of crowdsourcing and creativity support is a promising area of research~\cite{p22-kittur.pdf,Shneiderman2009}.
% dc2s22019,cc19workshop
and builds on 
%, allowing humans to excel in 
%the charactestics that 
%recombination, analogical transfer and divergent thinking while machines fall short in these fundamental characteristics needed for creativity.
a long line of work on augmenting human creativity and intellect
\cite{Doug_Engelbart-AugmentingHumanIntellect.pdf,GuilfordPresidentialAddress}.
More recently, research on supporting creativity has been sparking interest in the field of % Social Computing and
Human-Computer Interaction
% , where mixed-method studies on computational creativity support systems are gaining attention steadily
\cite{2019_chi-paper.pdf,dc2s22019} and design \cite{cc19workshop}. % p1235-frich.pdf, cc19workshop

\begin{marginfigure}% [20pc]%
  \begin{minipage}{\marginparwidth}%
    \centering%
    \includegraphics[trim={14cm 13cm 45cm 16cm},clip, width=\marginparwidth]{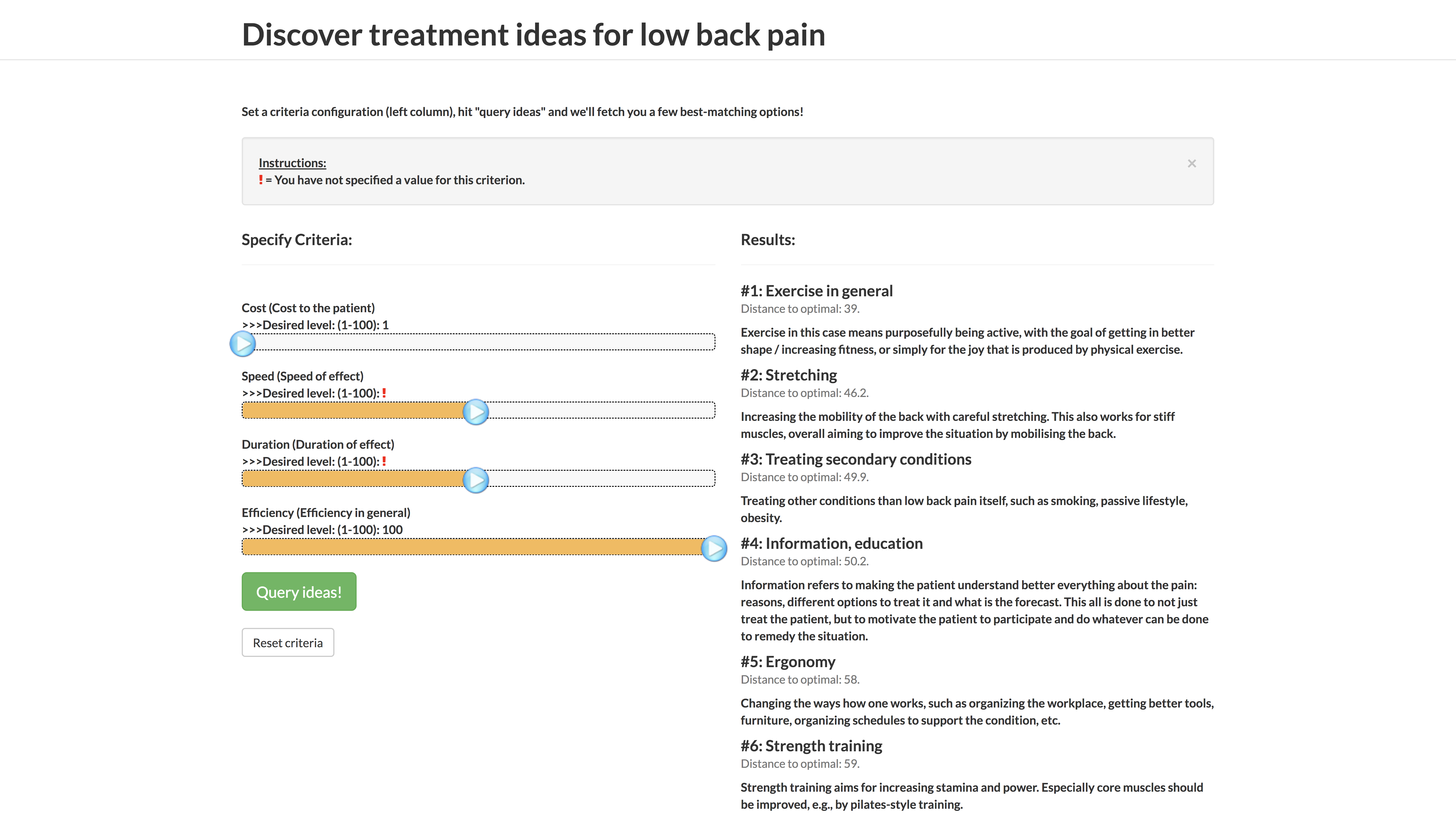}
    \caption{ArticleBot user interface for faceted filtering of ideas.}~\label{fig:articlebot}%
    \vspace{1em}%
  \end{minipage}%
\end{marginfigure}%

\begin{marginfigure}%[8pc]%
  \begin{minipage}{\marginparwidth}%
    \centering%
    \includegraphics[trim={3.5cm 10.5cm 16cm 4.5cm},clip, width=\marginparwidth]{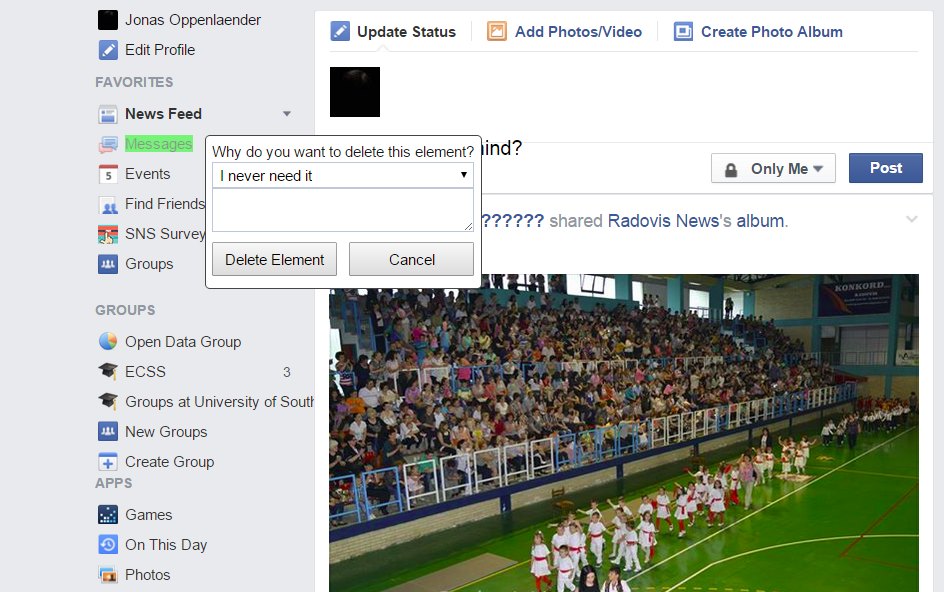}%
    \caption{CrowdUI user interface % dialog
    to provide justifications for manipulations of the website layout.}~\label{fig:crowdui}%
    \vspace{1em}
  \end{minipage}%
\end{marginfigure}%

\begin{marginfigure}%[-35pc]%
  \begin{minipage}{\marginparwidth}%
    \centering%
    \includegraphics[width=\textwidth]{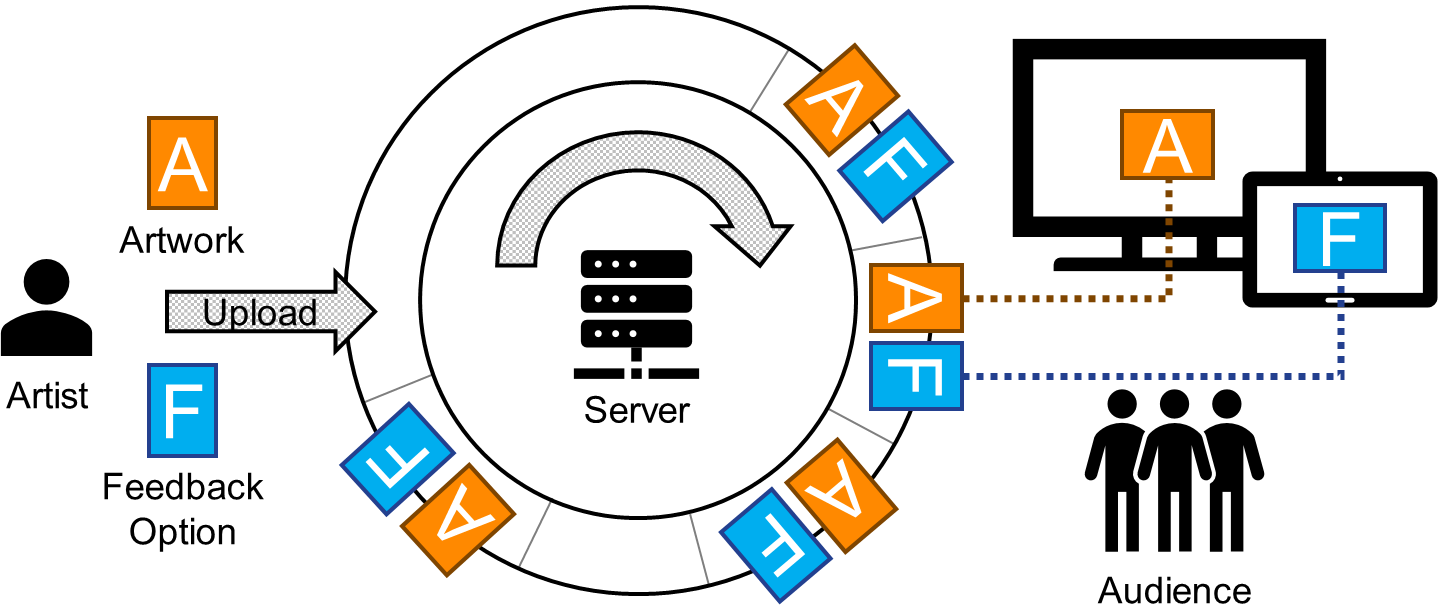}%
    \caption{High-level architecture of the situated feedback system.}~\label{fig:concept}%
  \end{minipage}%
\end{marginfigure}%

% Moreover, creativity is a social process. Studies from psychology show that groups of people with diverse backgrounds provide high quality ideas \cite{Hong16385}.
% %
% Using online deployments, it is becoming easier to reach out for diverse crowds, and therefore provide different contexts and backgrounds leading to diverse ideas~\cite{dennis2003}.
% For these reasons, crowdsourcing has been applied to a broad range of creative processes, from supporting research \cite{Vaish:2017} to helping unfold global-scale wicked problems~\cite{Cooper:2010,Cranshaw:2011}. Organizations have recognized the potential of crowds, with companies such as Innocentive, Quirky and OpenIDEO finding success in Open Innovation~\cite{hippel2011}. Given the inherent emphasis of crowdsourcing in collecting insights rapidly, inexpensively and accurately, it has been suggested as a key approach for creativity support~\cite{Andolina:2017,1214_yu.pdf}. 
%
% that goes back to the 1950s~\cite{GuilfordPresidentialAddress} and 1960s~\cite{Doug_Engelbart-AugmentingHumanIntellect.pdf}.

\textit{Crowd feedback systems}~\cite{luther-crowdcrit-cscw2015.pdf,p1433-xu.pdf} % crowdInnovationCourse-chi13.pdf, p25-redi.pdf,crowdmm2013.pdf
% \textit{Online feedback exchange systems}~\cite{p4454-foong.pdf}
are computer-mediated systems that
% allow for collecting
enable creative individuals to collect
feedback % and critique
from a large number of people online.
These systems provide an opportunity for asynchronously sourcing feedback, decision support and critique from a crowd with a diverse background. % and knowledge
% The crowd is heterogeneous and largely composed of people without specific subject matter expertise.

Researchers have investigated ways of increasing the perceived value of crowdsourced feedback~\cite{crowdInnovationCourse-hcomp2015.pdf,a063-krishna-kumaran.pdf},
% % % with better structure~\cite{luther-crowdcrit-cscw2015.pdf},
% % % rubrics~\cite{Posts_paper_3.pdf}, and
% % % scaffolding~\cite{mg_critiki_CaC2015_CameraReady.pdf}.
for instance by using rubrics to structure the feedback \cite{Posts_paper_3.pdf,luther-crowdcrit-cscw2015.pdf}
and interactive guidance to scaffold the feedback process \cite{paper55.pdf,mg_critiki_CaC2015_CameraReady.pdf}.
% Rubrics, for instance, can be used to structure the feedback and and guidance can be used to scaffold the feedback process which may % considerably
% improve the quality % and the perceived value
% of the crowdsourced feedback~\cite{Posts_paper_3.pdf,luther-crowdcrit-cscw2015.pdf,mg_critiki_CaC2015_CameraReady.pdf,paper55.pdf}.
% {Yuan et al.}, for instance, demonstrated that rubrics may increase the perceived value of crowdsourced design critiques for designers~\cite{Posts_paper_3.pdf}.
%
% \todo{LIST SOME MORE SPECIFIC WAYS, e.g. word cloud [specific citation]}
% Kang~et~al. enriched individual critique items with visual examples \cite{paper606.pdf}.
%
Using techniques such as the above,
crowdsourcing distributed % feedback % and design
critique
from novice crowds % was shown to potentially
may yield 
% more feedback than in existing online communities
feedback in comparable quality to expert critique \cite{luther-crowdcrit-cscw2015.pdf,Posts_paper_3.pdf}.

In the past year,
we explored the design space and limitations of crowd feedback systems 
% creativity support systems
% and the limitations of crowdsourced feedback
with three feedback systems that support
% We designed and built crowd-powered systems
% The systems support three
different types of creative work:
\begin{inparaenum}%[(a)]
  \item[] writing,
  \item[] web design, 
  \item[] and art critique.
\end{inparaenum}
% Note that all four submissions are currently still under review.
In the following sections, we give a brief overview of each system. For each system, we provide one idea for future research.

% ========================================
\section{ArticleBot: Supporting Exploratory Writing}
% ========================================

We designed and tested an intervention to support writers in finding and exploring different ideas and topics~\cite{ArticleBotINTERACT19}.
The lightweight system was adapted from Hosio~et~al.~\cite{Hosio:2016:LWC:3056355.3056393}.
The system provides decision support in form of short textual ideas that can be sorted according to different criteria (see Figure~\ref{fig:articlebot}).
Both the ideas, the criteria, and the rating of each idea across all criteria are provided by the crowd.
% {\color{red}
% We hypothesize that providing creative workers with an ordered list of ideas will help them in being more creative and in exploring the solution space.
% }

% We found that in a laboratory study, users (n=24) were strongly influenced by a status quo bias. Many used their own intuition to complete the given writing task, without turning to the creativity support system for help.
% Among the users that did use the system, we noticed that the readily-provided formative feedback cannibalised and stifled the participants' own divergent thinking.
% Many participants simply reproduced the top answer from the ordered list and completed the writing task without ever reflecting on the provided ideas.
% Consistent with the literature, we observed that
% feedback receivers pick the feedback that is well-aligned with their own mental model, and may disregard feedback that they perceive as being subjective \cite{p513-yen.pdf}.

Currently, the system collects all data up-front in a serendipitous fashion from visitors of a website or in a controlled fashion from crowd workers. While this database of ideas, criteria, and ratings holds value, it is expensive to create in terms of time taken and/or money spent and it is scoped to only one topic.
Future work could support writers with data collected from the crowd in real-time. To this end, instead of serendipitously collecting data organically from website visitors, the system could actively crowdsource information in real-time from online labour marketplaces, such as Mechanical Turk.

% ========================================
\section{CrowdUI: Supporting Web Design}
% ========================================

CrowdUI~\cite{CrowdUI} is a system that allows the community of a website to visually provide their feedback, using the website itself as a canvas.
Website visitors can directly manipulate (move, delete, resize) the elements of the webpage.
CrowdUI's multi-step process includes a tutorial to familiarise the user with the affordances of the system and a peer review of other users' creations.
The system allows designers to inspect the individual adaptations of the user interface. It also provides further decision support by visually aggregating user modifications of the website's user interface in heatmaps.
The system was evaluated in a remotely-administered user study with 48 users with promising results. % (n=48).

\begin{marginfigure}%[1.5pc]%
  \begin{minipage}{\marginparwidth}%
    \centering%
    \includegraphics[width=\marginparwidth]{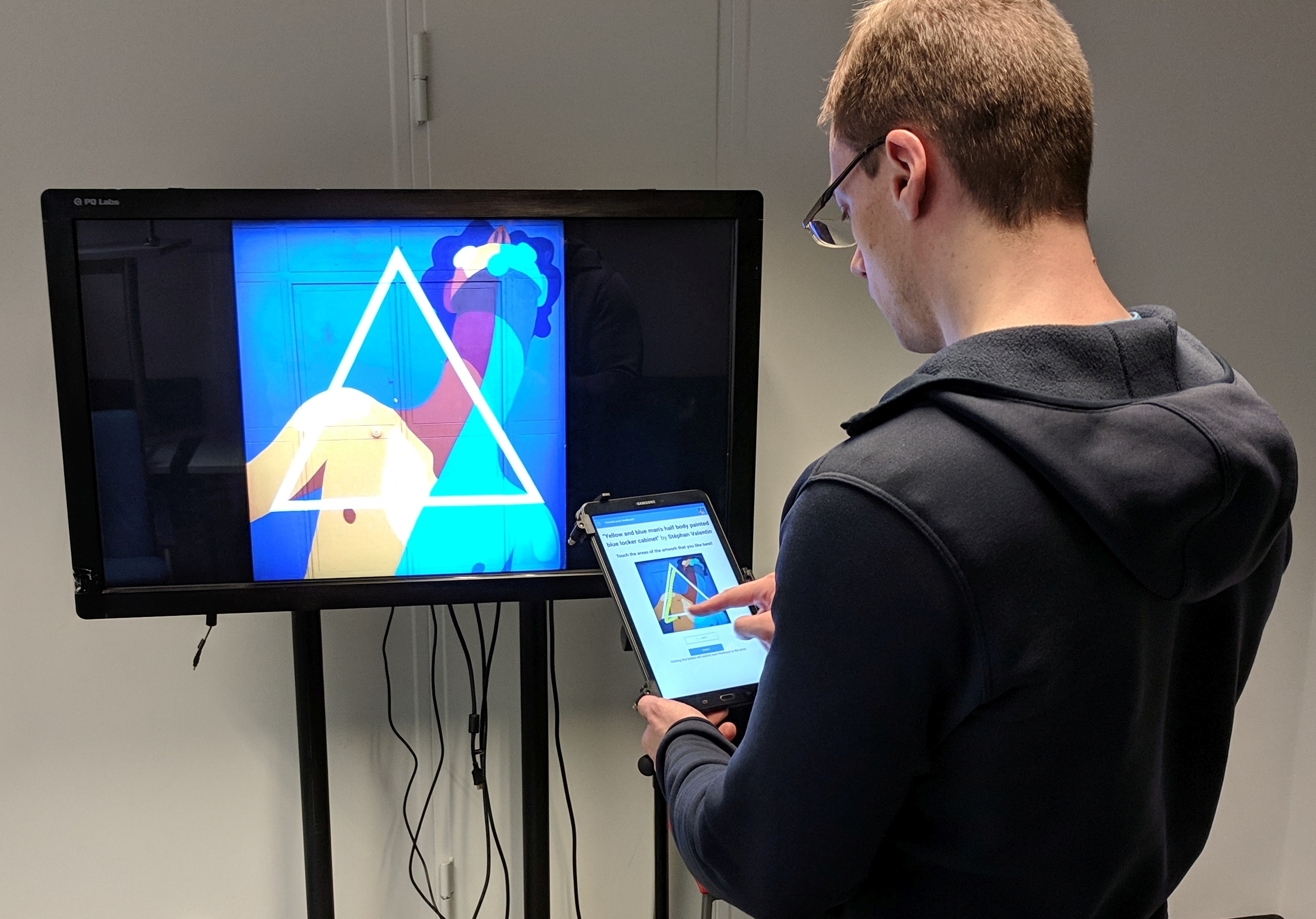}%
    \caption{Prototype of the situated feedback elicitation system. Artwork credit: Stéphan Valentin.}~\label{fig:simplex}%
    \vspace{1em}
  \end{minipage}%
\end{marginfigure}%

\begin{marginfigure}%[2pc]%
  \begin{minipage}{\marginparwidth}%
    \centering%
    \includegraphics[width=0.8\marginparwidth]{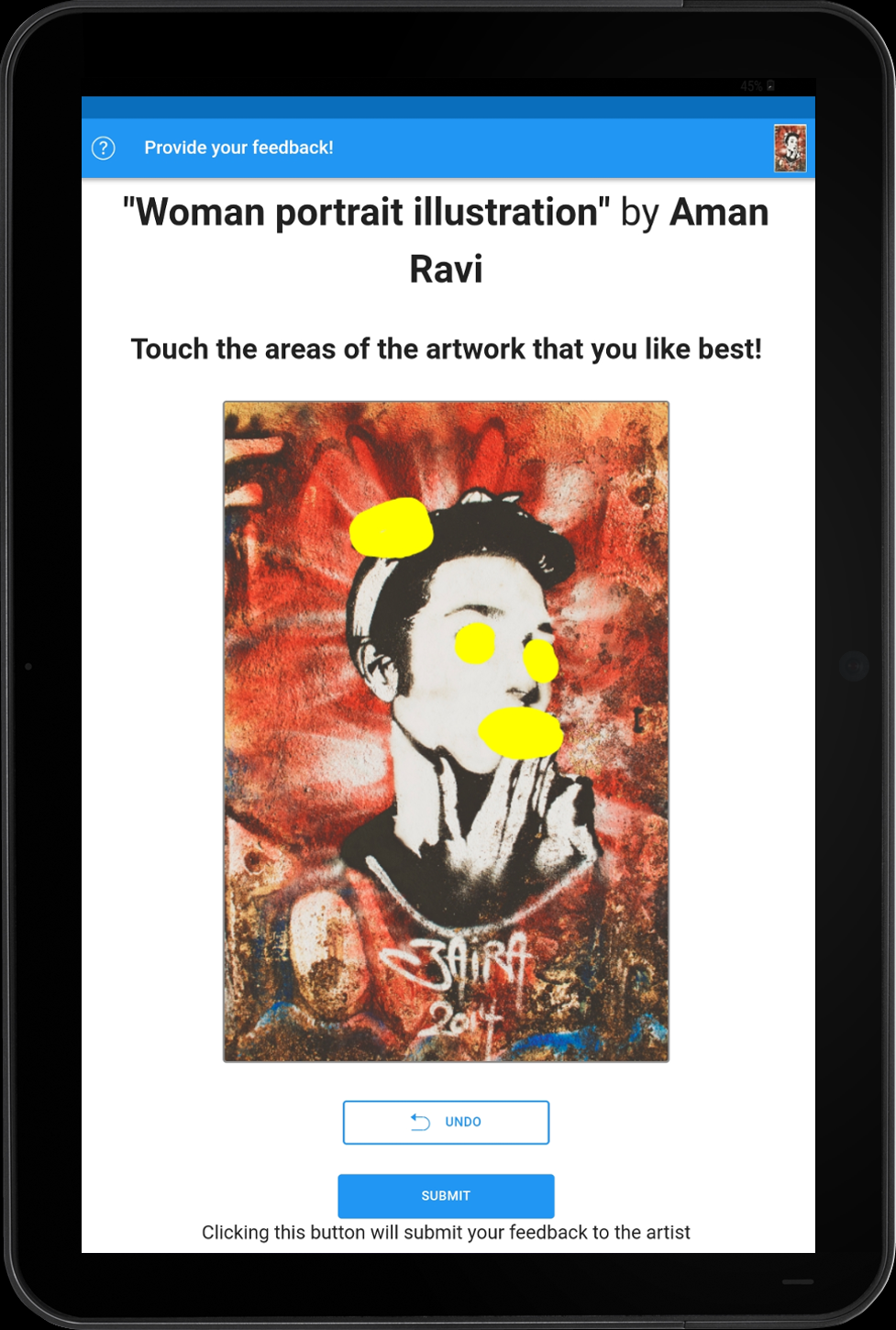}%
    \caption{Example of a type of feedback in the situated feedback system. Artwork credit: Aman Ravi.}~\label{fig:simplexfeedback}%
  \end{minipage}%
\end{marginfigure}%

% Web design is a process that involves many design decisions.
Web designers and developers are faced with many decisions during the website design process.
% Future work could support web designers and web developers in making decisions in the design process.
One idea for future work involves crowdsourcing preference judgements from the crowd in a web-based game with a purpose (GWAP) to help the developer in making informed decisions and to explore the space of possible design alternatives.
The game would automatically tweak variables in the design space (\textit{e.g.} changing the font style of a headline to bold).
It would then elicit a pairwise comparison of the design alternatives from its players, resulting in a ranking of individual attributes.
Similar web-based games, for instance \textit{Can't Unsee} (\href{https://cantunsee.space}{https://cantunsee.space}),  have been found to be very engaging.

\section{SIMPLEX: Eliciting Art Critique}

In our third system~\cite{Simplex}, we explored situated crowdsourcing % \cite{situatedcrowdsourcing}
and public displays for supporting artists with summative feedback.
% we turned to exploring situated technology, as it allows us to study and observe the usage of the system in person.
% With our conceptual architecture for the elicitation of art critique (see Figure~\ref{fig:concept}), we explore how artists can be supported with situated feedback and technology.
The system enables the situated crowd to provide feedback to artists via an installation that consists of two public displays (see Figures~\ref{fig:concept} and \ref{fig:simplex}).
A digital artwork is displayed on the main
screen of the installation. Feedback is given on the smaller
touch screen positioned in front of the main screen. Once feedback is given, a new artwork is displayed on the main screen.
In a needfinding study with two artists and a user study with 12 participants, we evaluated eight different types of feedback (see the example in Figure~\ref{fig:simplexfeedback}).

Future work will evaluate the ecological feasibility of the SIMPLEX crowd feedback system in a longitudinal field study.
The system will be evaluated using a mixed-method approach, combining the quantitative analysis of interaction data with qualitative insights from semi-structured interviews.
Further evaluation could also include the application of the system in a specific use case, \textit{e.g.} in the context of a design-oriented University course to evaluate architectural designs from different perspectives.

\section{Conclusion}
% -------------------

In this position paper, we presented three crowd feedback systems that support creative work.
Our exploration of different modes and types of feedback highlighted that there are many opportunities to support creative work with crowd feedback, beyond mere text-based feedback.

Our ongoing work in the form of a structured literature review
% \cite{525444systematicreviewsguide.pdf}
% survey the current strategies of creative individuals for 
will guide the development of a taxonomy of activities that creative individuals engage in when they evaluate feedback from the crowd.
This work will culminate in a conceptual framework to highlight critical processes that affect the sensemaking of crowdsourced feedback.
We envision this design framework to create a basis for discussing challenges and issues related to the design of interactive feedback elicitation systems.

\bibliographystyle{ACM-Reference-Format}
\bibliography{submission}

\end{document}